\documentclass[usenatbib]{mnras}
\usepackage[T1]{fontenc}
\usepackage{ae,aecompl}
\usepackage{bm}
\usepackage{amsmath,amssymb}
\usepackage{dcolumn}
\usepackage[dvips]{graphicx}
\usepackage{epsf}
\usepackage{color}
\usepackage{epsf}
\usepackage{epsfig}
\usepackage{graphicx}
\usepackage{longtable}
\usepackage{psfrag}
\usepackage{txfonts}
\usepackage[Symbol]{upgreek}       
\usepackage{ulem}
\usepackage[usenames,dvipsnames]{xcolor}

\def \be {\begin{equation}} 
\def \ee {\end{equation}} 
\def \bea {\begin{eqnarray}} 
\def \eea {\end{eqnarray}}

\begin{document}

\title[The scale of homogeneity with SDSS-IV DR14 quasars]{Measuring the scale of cosmic homogeneity with SDSS-IV DR14 quasars}

\author[R. S. Gon\c{c}alves et al.]{
 \parbox{\textwidth}{
  R. S. Gon\c{c}alves $^{1,2}$\thanks{E-mail: \texttt{rsousa@on.br}},
  G. C. Carvalho $^{2,3}$, 
  C. A. P. Bengaly $^{4}$, 
  J. C. Carvalho $^{2}$, 
  J. S. Alcaniz$^{2,5}$
 }
 \vspace{0.4cm}\\
 \parbox{\textwidth}{
  $^{1}$Departamento de F\'isica, Universidade Federal do Maranh\~ao, 65080-805, S\~ao Lu\'is, Maranh\~ao, Brasil\\
  $^{2}$Observat\'orio Nacional, 20921-400, Rio de Janeiro - RJ, Brasil\\
  $^{3}$Departamento de Astronomia, Universidade de S\~ao Paulo, 05508-090, S\~ao Paulo - SP, Brasil\\
  $^{4}$Department of Physics \& Astronomy, University of the Western Cape, Cape Town 7535, South Africa\\
  $^{5}$Departamento de F\'{\i}sica, Universidade Federal do Rio Grande do Norte, 59072-970, Natal - RN, Brasil\\
 }
}

\pagerange{\pageref{firstpage}--\pageref{lastpage}}


\maketitle

\label{firstpage}

\begin{abstract}
The quasar sample of the fourteenth data release of the Sloan Digital Sky Survey (SDSS-IV DR14) is used to determine the cosmic homogeneity scale in the redshift range $0.80<z<2.24$.  We divide the sample into 4 redshift bins, each one with $N_{\rm q} \geq 19,000$ quasars, spanning the whole redshift coverage of the survey and use two correlation function estimators to measure the scaled counts-in-spheres and its logarithmic derivative, i.e., the fractal correlation dimension, $D_2$. Using the $\Lambda$CDM cosmology as the fiducial model, we first estimate the redshift evolution of quasar bias and then the homogeneity scale of the underlying matter distribution $r_{\rm{hom}}^{\rm{m}}$. We find that $r_{\rm{hom}}^{\rm{m}}$ exhibits a decreasing trend with redshift and that the values obtained are in good agreement with the $\Lambda$CDM prediction over the entire redshift interval studied. We, therefore,  conclude that the large-scale homogeneity assumption is consistent with the largest spatial distribution of quasars currently available.        
\end{abstract}



\section{Introduction}

Given the rapidly increasing amount of cosmological data, not only it is possible to test the standard cosmological model (SCM), i.e., the $\Lambda$CDM cosmology, with unprecedented precision~\citep{planck16a}, but also to assess the validity of its fundamental pillars in light of these observations. Among these pillars, the Cosmological Principle (CP) states that the Universe is spatially isotropic and homogeneous on large scales, which means that cosmological distances and ages can be well described by the Friedmann-Lema\^itre-Robertson-Walker (FLRW) metric. Hence, one of the greatest challenges of the SCM is to determine whether these assumptions actually hold true. If one proves that this is not the case, then a profound reinterpretation of our understanding of the Universe is needed (see, e.g.,~\citealt{ellis}). 

Cosmological isotropy has been directly tested using different cosmological observations, such as CMB~\citep{schwarz16, planck16b}, galaxy and radio source counts~\citep{alonso15, colin17, bengaly18a, bengaly18b, rameez18}, weak lensing convergence~\citep{marques18}, and cosmic expansion probed by Type Ia Supernova distances~\citep{lin16, andrade18}. Homogeneity, on the other hand, is much harder to probe because we can only observe down the past lightcone, not directly on time-constant hypersurfaces~\citep{clarksonmaartens10, maartens11, clarkson12}. Nonetheless, one can test the consistency between the FLRW hypothesis and observations performed in the past lightcone. Any strong disagreement can in principle be ascribed to a departure of CP. Presently, most of the analyses performed found a transition scale from a locally clumpy to a smooth, statistically homogeneous Universe, using galaxy or quasar counts in the interval $70 < r_{\rm hom} < 150 \; \mathrm{Mpc/h}$~\citep{hogg05, yadav10, scrimgeour12, nadathur13, alonso15, sarkar16, laurent16, ntelis17, goncalves18}, yet some authors dispute these results~\citep{syloslabini09, syloslabini11, park17}.   

In this paper we estimate the cosmic homogeneity scale using the spatial distribution of quasar number counts from the recently released quasar sample (DR14) of the Sloan Digital Sky Survey IV (SDSS-IV)~\citep{sdss17}. This data set covers a very large redshift interval, $0.80 < z < 2.25$, hence providing the largest volume ever in which this kind of analysis has been carried out. We divide the sample into 4 redshift bins spanning the whole redshift coverage of the survey and use two correlation function estimators to study whether there is clear evidence for such homogeneity scale in all ranges. Our results of scaled counts-in-sphere (see Sec. 3) are used to estimate a possible redshift dependence of quasar bias. We then use these results to compare our measurements of the homogeneity scale with the transition scale for the underlying matter distribution predicted by the SCM scenario since any inconsistency found may hint at a failure of the FLRW assumption on describing the observed Universe. We find a good agreement with previous results using different types of tracers of large-scale structure. This paper is organised as follows. In Section 2 we briefly describe the observational data used in our analysis. Section 3 discusses the method and correlation function estimators adopted. The results and discussions are presented in Section 4. Section 5 summarises our main conclusions.

\section{Observational data}

The quasar data release fourteen (DR14) of the extended Baryonic Oscillation Sky Survey (eBOSS)~\citep{eboss16} covers a total effective area of about $2,000$ deg$^2$  comprising $~150,000$ quasars distributed in the redshift interval of $0.80 < z < 2.25$.  As in previous BOSS data releases, DR14 is divided into two different regions in the sky, named north and south galactic caps. Here we are interested in exploring the homogeneity transition at redshifts $z > 0.8$, and we use only the north galactic cap. We split the sample into four redshift bins, as presented in Table~\ref{table1}, of width $0.26 < \Delta z < 0.37$. The mean redshift of the bins are $\bar{z} = 0.985, 1.35, 1.690, 2.075$. As shown in Table~\ref{table1}, the number of quasars in each redshift bin is $N_{\rm q} \geq 19,000$, thus providing good statistical performance for the analysis. In order to avoid correlation between neighbouring bins, we carefully choose non-contiguous redshift bins.

\section{Methodology}

In this section we present the method used to measure the transition to homogeneity in the distribution of the SDSS-IV DR14 quasar catalogue. 

In order to explore the spatial homogeneity of the quasar sample, we use the so-called scaled counts-in-spheres, $\cal N$, that is defined as the ratio between the average number of points within the observational catalogue and the average number of points within random catalogues~\citep{scrimgeour12, laurent16}. We use twenty random catalogues that are generated by a  Poisson  distribution  with  the  same  geometry  and  completeness  as  the DR14.

In order to calculate $\cal N$, an important quantity is the 3D separation distance between two different objects ($\rho$) within the sample, which can be calculated according to
\begin{equation}
\label{eq:r}
\rho=\sqrt{d(z_1)^2 + d(z_2)^2 - 2d(z_1)d(z_2)\cos{\theta}} \;,
\end{equation}
where 
$\cos{\theta} = \sin{\delta_1}\sin{\delta_2} + \cos{\delta_1}\cos{\delta_2}\cos{(\alpha_1 - \alpha_2)}$ and 
${\alpha_i}$ and ${\delta_i}$ ($i=1,2$) are, respectively, their right ascension and declination. The radial comoving distance $d(z)$ is defined as
\begin{equation}
\label{eq:dz}
d(z) = \int_0^z cdz'/H(z'); \; H(z) = H_0\sqrt{\Omega_{\rm m}(1+z)^3 + \Omega_\Lambda} \;.  
\end{equation}
with our fiducial cosmology fixed at $\Omega_{\rm m} = 0.313$, $\Omega_{\Lambda} = 0.687$ and $H_0 = 67.48 \, \mathrm{km/s/Mpc}$~\citep{planck16a}. 

From $\cal N$, one can compute the fractal correlation dimension $D_2$, for a given distance $\rho = r$, defined by 
\begin{equation}
\label{defD2}
D_2(r) \equiv {d\ln {\mathcal{N}(<r)} \over d\ln\!r} + 3 \;.
\end{equation}
As the distribution approaches the homogeneity scale, the scaled counts-in-spheres is ${\cal N} \rightarrow 1$, and hence ${D_2} \rightarrow 3$~\citep{scrimgeour12}.

Several estimators for $\cal N$ have been explored in the literature~\citep{goncalves18}. Basically there are two classes of them: the first one is based on the average distribution of points in the sky~\citep{scrimgeour12} whereas the second is based on correlation functions~\citep{laurent16, ntelis17}. We use the latter in this paper, as explained below.

\subsection{Peebles-Hauser}

The Peebles-Hauser (PH) estimator is based on the correlation function proposed by \cite{peebleshauser74} and can be expressed as
\begin{equation}
\label{NPH}
{\cal N}(\!<\! r)\! \equiv \!
\frac{\sum^{r}_{\rho = 0} DD(\rho)}{\sum^{r}_{\rho = 0} RR(\rho)} \;,
\end{equation}
where $DD(\rho)$ is defined as the pair of observed quasars counts within a separation radius $\rho$ normalised by the the total number of pairs, $N^{\rm obs}(N^{\rm obs} - 1)/2$ and $RR(\rho)$ follows the same definition for the random catalogues. 

\subsection{Landy-Szalay}

The Landy-Szalay (LS) estimator is based on the correlation function proposed by \cite{landyszalay93}. In addition to the quantities $DD(\rho)$ and $RR(\rho)$ defined above, $DR(\rho)$ denotes the pair of quasar-random counts normalised by the available number of pairs. Hence, the LS estimator is given by
\begin{equation}
\label{NLS}
{\cal N}(\!<\!r) \equiv 
1 + \frac{\sum^{r}_{\rho = 0} [DD(\rho)-2DR(\rho)+RR(\rho)]}{\sum^{r}_{\rho = 0} RR(\rho)} \;.
\end{equation}
For the PH and LS estimators, the quantity $D_2(r)$ is calculated by replacing Eqs.~(\ref{NPH}) and (\ref{NLS}) into Eq.~(\ref{defD2}), respectively.

\begin{table}
\centering
\begin{tabular}{|c|c|c|c|}
\hline
\,\,$z$ interval \,\,& \,\,$\bar{z}$ \,\,& 
\,\,$N_{\rm q}$\,\, \\
\hline \hline
\,\,0.80-1.17 \,\,& 0.985 \,\,& 
19163 \\
\,\,1.22-1.48 \,\,& 1.350 \,\,& 
19104 \\
\,\,1.56-1.82 \,\,& 1.690 \,\,& 
19141 \\
\,\,1.90-2.25 \,\,& 2.075 \,\,& 
19225 \\
\hline
\end{tabular}
\caption{The four redshift bins used in the analysis and their numbers: redshift range ($z$), mean redshift ($\bar{z}$) and number of quasars ($N_{\rm q}$).}
\label{table1}
\end{table}

\section{The homogeneity scale}

In order to estimate the homogeneity scale from the SDSS-IV DR14 quasars (henceforth $r_{\rm hom}^{\rm q}$) we first obtain the values of $D_2(r)$ for each redshift slice using both estimators above. Then, a polynomial fitting is used to obtain the corresponding scale where the transition to homogeneity is identified (we refer the reader to \cite{goncalves18} for more details). 

Following  previous  analyses~\citep{scrimgeour12, laurent16, ntelis17}, we define the homogeneity scale as the characteristic scale where the Universe can be considered homogeneous within 1\% of the expected fractal dimension for a homogeneous distribution, $D_2=3$ - thus, the scale where the spatial distribution of QSOs reaches $D_2=2.97$. Although arbitrary, the 1\%-criterion is widely used in the literature, and is justified given some observational issues, such as the survey geometry and incompleteness, as well as the sample noise. Since we use twenty random catalogues to obtain $D_2$, we also perform a bootstrap analysis over the twenty values of $r_{\rm hom}^{\rm q}$, quoting its mean and standard deviation as our measurement of the homogeneity scale and its corresponding uncertainty, respectively\footnote{For other works using the bootstrap method to estimate clustering uncertainties, see~\cite{norberg09,ansarinejad18}}. 

\begin{figure*}
\vspace{-0.1cm}
\centering
\includegraphics[scale=0.51]{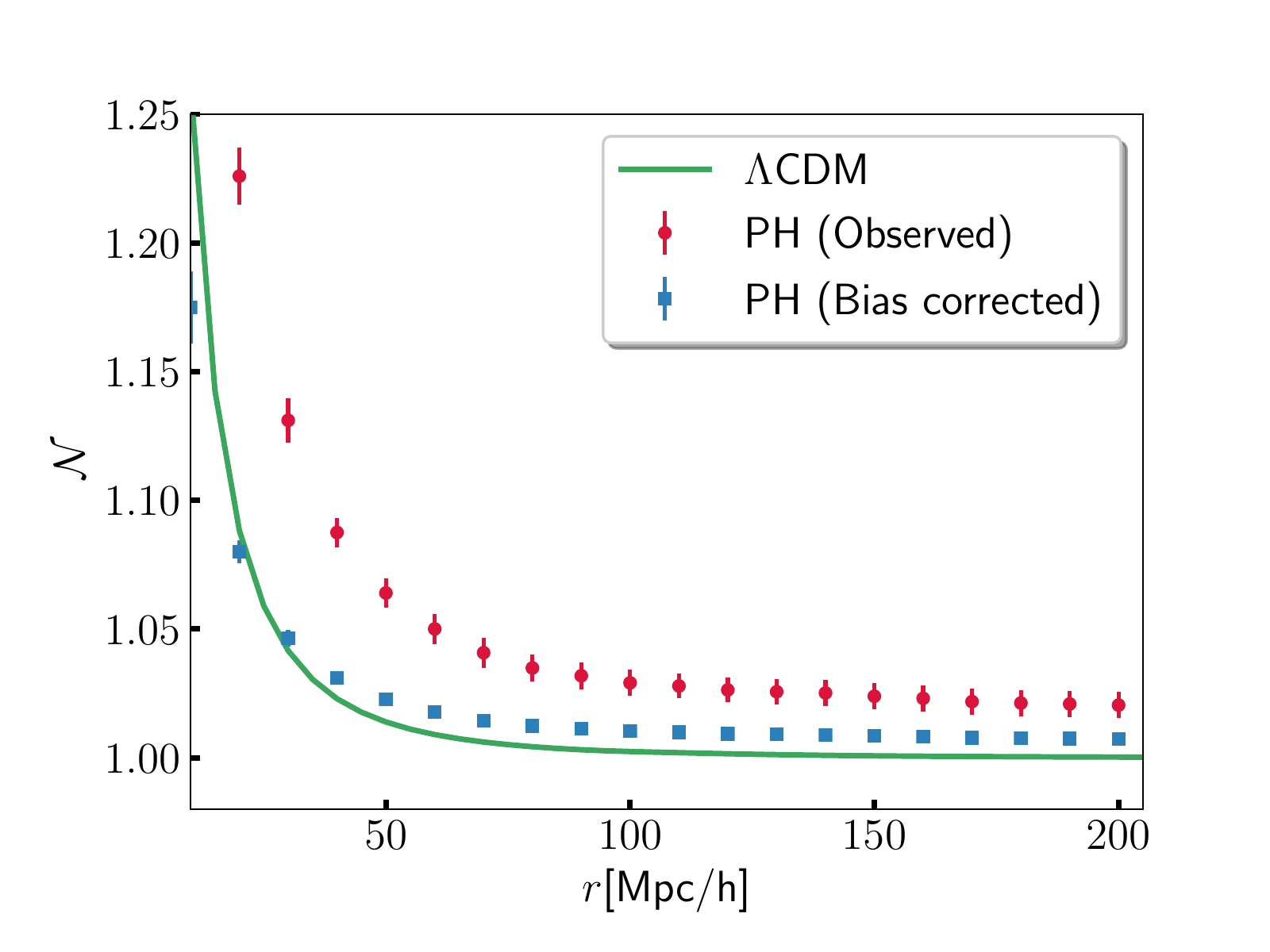}
\includegraphics[scale=0.51]{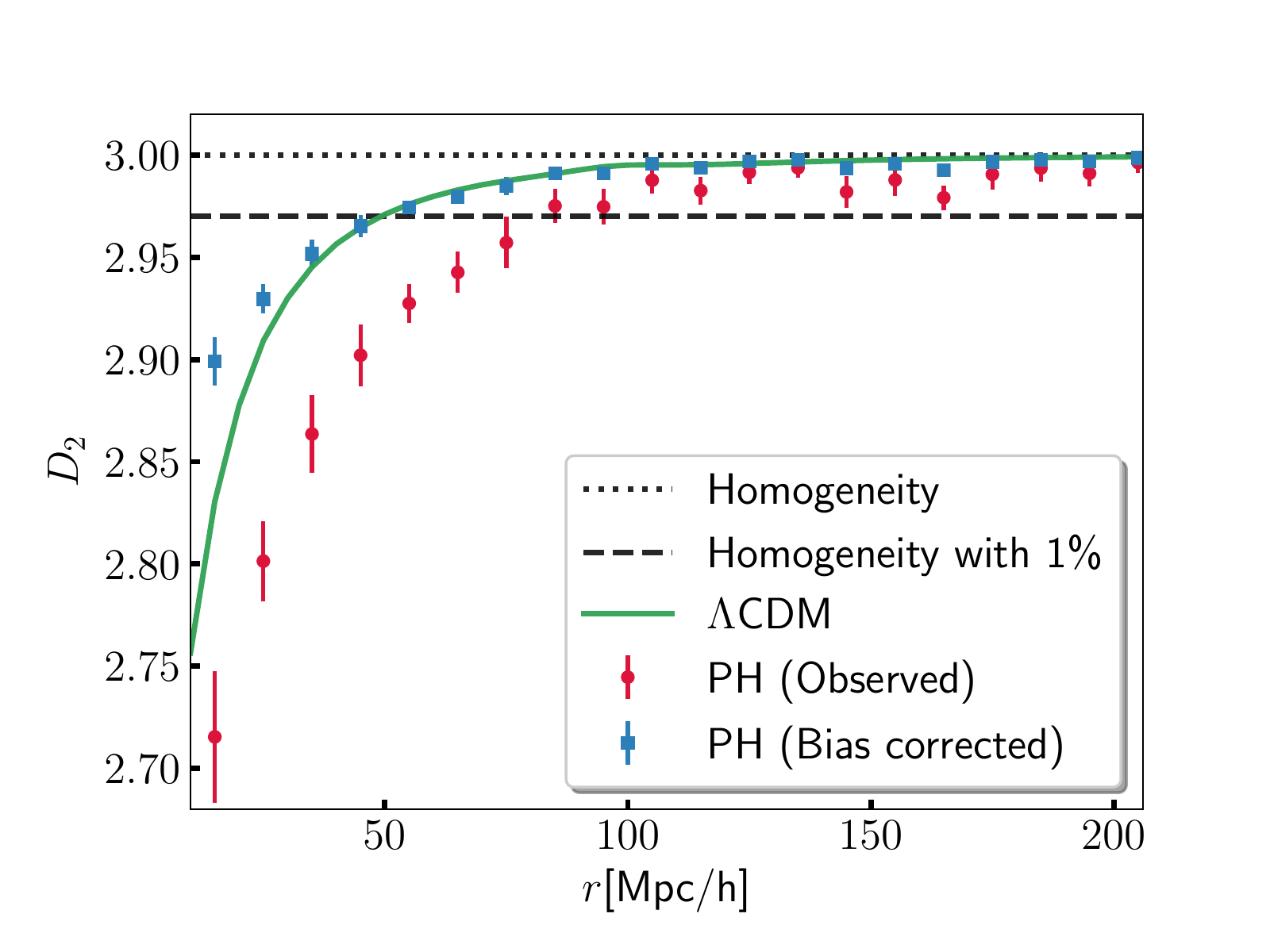}
\caption{{\it{Left:}} Scaled counts-in-spheres $\mathcal{N}(<r)$ obtained for the redshift bin $0.80<z<1.17$ ($\bar{z} = 0.985$) assuming the PH estimator. The red and green points correspond, respectively, to the observed and bias-corrected values of $\mathcal{N}(<r)$ assuming the best-fit value of $b_{\rm q}(z)$. {\it{Right:}} Correlation dimension as a function of the scale $r$ for the PH estimator at $\bar{z} = 0.985$. The points represent observational results whereas the dotted and dashed horizontal lines indicate $D_2=2.97$ and $D_2=3$, respectively.}
\label{fig1}
\end{figure*}

\begin{table}
\centering
\begin{tabular}{|c|c|c|c|}
\hline
$\bar{z}$ \,\,& \,\, $r_{\rm{hom}}^{\rm{q}}$ \,\,& \,\, $r_{\rm{hom}}^{\rm{m}}$ \,\,& \,\, $r_{\rm{hom}}^{\rm{th}}$ \,\, \\
\hline
\,\,0.985 \,\,&  82.96 $\pm$ 8.30 \,\,&\,\, 48.78 $\pm$ 3.82 \,\,&\,\, 52 \\
\,\,1.350 \,\,& 101.55 $\pm$ 7.51 \,\,&\,\, 40.56 $\pm$ 3.39 \,\,&\,\, 45 \\
\,\,1.690 \,\,&  85.87 $\pm$ 8.75 \,\,&\,\, 36.19 $\pm$ 3.45 \,\,&\,\, 40 \\
\,\,2.075 \,\,& 102.55 $\pm$ 6.73 \,\,&\,\, 27.91 $\pm$ 3.91 \,\,&\,\, 34 \\
\hline
\end{tabular}
\caption{Homogeneity scale obtained from the estimator PH in each redshift slice. The second column shows the $r_{\rm hom}^{\rm{q}}$ values obtained from the real data; the third  column corresponds to the homogeneity scale after accounting for the bias as described in Section 4.1 ($r_{\rm{hom}}^{\rm{m}}$); the fourth column shows the theoretical prediction of our fiducial model ($r_{\rm{hom}}^{\rm{th}}$). All these values are given in units of Mpc/h.}
\label{tableresultPH}
\end{table}

\begin{table}
\centering
\begin{tabular}{|c|c|c|c|}
\hline
$\bar{z}$ \,\,& \,\, $r_{\rm{hom}}^{\rm{q}}$ \,\,& \,\, $r_{\rm{hom}}^{\rm{m}}$ \,\,& \,\, $r_{\rm{hom}}^{\rm{th}}$ \,\, \\
\hline
\,\,0.985 \,\,& 83.65 $\pm$  8.82 \,\,&\,\, 52.93 $\pm$ 7.55 \,\,&\,\, 52 \\
\,\,1.350 \,\,& 93.67 $\pm$ 15.73 \,\,&\,\, 40.43 $\pm$ 5.64 \,\,&\,\, 45 \\
\,\,1.690 \,\,& 78.28 $\pm$ 10.21 \,\,&\,\, 36.66 $\pm$ 4.80 \,\,&\,\, 40 \\
\,\,2.075 \,\,& 95.65 $\pm$ 19.35 \,\,&\,\, 29.94 $\pm$ 3.35 \,\,&\,\, 34 \\
\hline
\end{tabular}
\caption{The same as in Table~\ref{tableresultPH}, but for the LS estimator instead.}
\label{tableresultLS}
\end{table}

Our first measurements of $r_{\rm hom}^{\rm{q}}$ are presented in the second column of Tables~\ref{tableresultPH} and~\ref{tableresultLS}. For the sake of example, Fig.~\ref{fig1} shows the results for both ${\cal N}$ and $D_2$ obtained at the redshift $\bar{z} = 0.985$ using the PH estimator (similar results are also obtained for the LS estimator). The red points stand for the observational data with the respective errors. In the right panel of Fig.~\ref{fig1}, the horizontal lines correspond to the $D_2 = 2.97$ and $D_2 = 3$ thresholds. From our analysis we find that the PH estimator provides smaller error bars than the LS estimator. A possible explanation may be attributed to the fact that our approach does not use mock catalogues. Instead, we relied on an accurate bootstrap analysis of the twenty random catalogues we have produced. A full assessment of the homogeneity scale using mock catalogues that reflects the matter power spectrum, redshift distribution, QSO clustering, and angular selection function of the surveys, are in progress (Goncalves et al. in prep.).


\subsection{The quasar bias}

In order to estimate the homogeneity scale for the underlying matter distribution, a correction due to the quasar bias ($b_{\rm q}$) needs to be introduced. This quantity can be estimated from the measurements of ${\cal{N}}$ as follows. First, we compute the two-point correlation function from the matter power spectrum, $P_{\rm m}(\kappa,\bar{z})$, i.e.,
\begin{equation}
\label{TheoreticalPS}
\xi(s,\bar{z}) = \frac{1}{2\pi^2} \int P_{\rm m} (\kappa,\bar{z})\frac{\sin(\kappa s)}{\kappa s} \kappa^2 d\kappa \; ,
\end{equation}
where $P_{\rm m}(\kappa,\bar{z})$ is obtained from the CAMB code~\citep{lewis00} assuming the same fiducial model described earlier. We then calculate the theoretical $\mathcal{N}(r)_{\rm th}$ from
\begin{equation}
\label{TheoreticalCursN}
{\cal N}_{\rm th} (< r,\bar{z}) = \frac{3}{4\pi r^3} \int_ {0}^{r} (1+\xi(s,\bar{z})) 4 \pi s^2 ds \; .
\end{equation}
Finally, the observational value of the correlation integral for the matter distribution, ${\cal N}_{\rm m}(< r,\bar{z})$, is related to the observed one, ${\cal N}(< r,\bar{z})$, by means of (see \cite{laurent16} for a discussion)
\begin{equation}
\label{RelationCursN}
{\cal N}_{\rm m}(< r,\bar{z}) = \frac{{\cal N}(< r,\bar{z}) - 1}{b_{\rm q}^2} + 1  \;,
\end{equation}
in which the quantity $b_{\rm q}$ is estimated through a $\chi^2$ minimisation procedure, i.e., 
$\chi^2 = \sum_{\rm i} {({\cal N}^{\rm i}_{\rm m} - {\cal N}^{\rm i}_{\rm th})^2}/{\sigma_{{\cal N}_{\rm m}^{\rm i}}^2}$, 
where $i$ corresponds to the $i$-th data point of ${\cal N}_{\rm m}$ in Fig.~\ref{fig1}, and $\sigma_{{\cal N}_{\rm m}}$ denotes its respective uncertainty, at each $\bar{z}$. This procedure is performed for ${\cal N}^{\rm i}$ obtained from both PH and LS estimators. Note that the bias may vary with the scale $r$. Here, however, we assume it to be constant along the redshift bin (see, e.g., Sec. 5.2 of \cite{laurent16} for a discussion). It is worth noticing that we do not make use of the full covariance matrix of the N (< r) in this analysis. However, we do not expect that it would change significantly the accuracy of the present results. A more thorough estimate of the clustering bias will be pursued in a future work.

\begin{table}
\centering
\begin{tabular}{|c|c|c|}
\hline
$\bar{z}$ \,\,& \,\,$b_{\rm q} \pm \sigma_{b_{\rm q}} \; \mathrm{(PH)}$\,\,& \,\,$b_{\rm q} \pm \sigma_{b_{\rm q}} \; \mathrm{(LS)}$\,\, \\
\hline
\,\,0.985 \,\,& 1.68 $\pm$ 0.020 \,\,&\,\, 1.61 $\pm$ 0.025 \\
\,\,1.350 \,\,& 2.10 $\pm$ 0.025 \,\,&\,\, 2.02 $\pm$ 0.030 \\
\,\,1.690 \,\,& 2.28 $\pm$ 0.030 \,\,&\,\, 2.27 $\pm$ 0.035 \\
\,\,2.075 \,\,& 2.93 $\pm$ 0.025 \,\,&\,\, 2.86 $\pm$ 0.035 \\
\hline
\end{tabular}
\caption{The quasar bias obtained for each redshift bin using PH and LS estimators. To obtain these results we assume that ${b_{\rm q}}$ has no dependence on the scale $r$.}
\label{table4}
\end{table}

In order to model such dependence with $z$, we explore three possible parameterisations of $b_{\rm q}(z)$: $b_{\rm q}(z) = b_0 + b_1(1+z)$ $(\rm{P1})$, $b_{\rm q}(z) = b_0 + b_1(1+z)^2$ $(\rm{P2})$ and $b_{\rm q}(z) = b_0 + b_1z(1+z)/(1+z^2)$ $(\rm{P3})$. For completeness, we also consider a constant parameterisation $b_{\rm q} = b_0$ $(\rm{P4})$. The data points of Table~(\ref{table4}) are used to estimate the free parameters of  the above parameterisations through a standard $\chi^2$ minimisation. By means of the Bayesian information criterion (BIC)~\citep{schwarz78,Liddle:2007fy}, we select P2 as the best parameterisation of $b_{\rm q}(z)$ since the others provide $\Delta {\rm{BIC}} > 20$ with respect to it. At 1$\sigma$ level, the best-fit values for P2 read 
$$\label{bz} 
\mathrm{\bf PH:} \; b_{\rm q}(z) = 0.82 (\pm 0.02) + 0.22 (\pm 0.01) (1+z)^2\;,
$$
$$
\mathrm{\bf LS:} \; b_{\rm q}(z) = 0.75 (\pm 0.02) + 0.22 (\pm 0.01)(1+z)^2\;,
$$ 
whose $b_{\rm q}$ values at each $\bar{z}$ are given in Table~\ref{table4}. We note that the results above are in good agreement with previous bias estimates reported in the literature. For instance, previous analyses using the SDSS (BOSS) DR9 and DR12 quasar samples obtained, respectively, $b_{\rm q} = 3.80 \pm 0.30$~\citep{white12} and $b_{\rm q} = 3.54 \pm 0.11$~\citep{efterkharzadeh15}, both at $\bar{z} = 2.39$. Our best-fit for P2 provides $b_{\rm q}(z = 2.39) = 3.34 \pm 0.13$ (PH) and $b_{\rm q}(z = 2.39) = 3.28\pm 0.13$ (LS), thus compatible with the previous results within $2\sigma$ (CL). 

\subsection{Bias-corrected measurements}

In order to compare our results with the SCM prediction, we have to correct the fractal correlation dimension due to the quasar bias. We do so by obtaining the homogeneity scales at which $D_2 \rightarrow 2.97$ from the bias corrected correlation integral, ${\cal N}_{\rm m}(< r,\bar{z})$, after we input the bias best-fitted values shown in Table~\ref{table4} in Eq.~\ref{RelationCursN}. Then, we compare with the theoretically expected result using $D_2$ as obtained from Eq.~\ref{TheoreticalCursN}. These values will be hereafter denoted by $r_{\rm hom}^{\rm m}$ and $r_{\rm th}^{\rm m}$, respectively. 

We find that the transition to homogeneity is reached earlier for the matter distribution than the quasars due to the effect of $b_{\rm q}(z)$. This can be seen by looking back to the right panel of Fig.~\ref{fig1}, while we clearly note that blue curve, which represents the bias-corrected fractal dimension, reaches $2.97$ before the red one, which represents the fractal dimension observed with quasars. We also provide the result from the fiducial cosmology model in the green solid line of the same plot, where we note that it is fully consistent with the bias-corrected measurements. 

\begin{figure}
\vspace{-0.1 cm}
\centering
\includegraphics[scale=0.55]{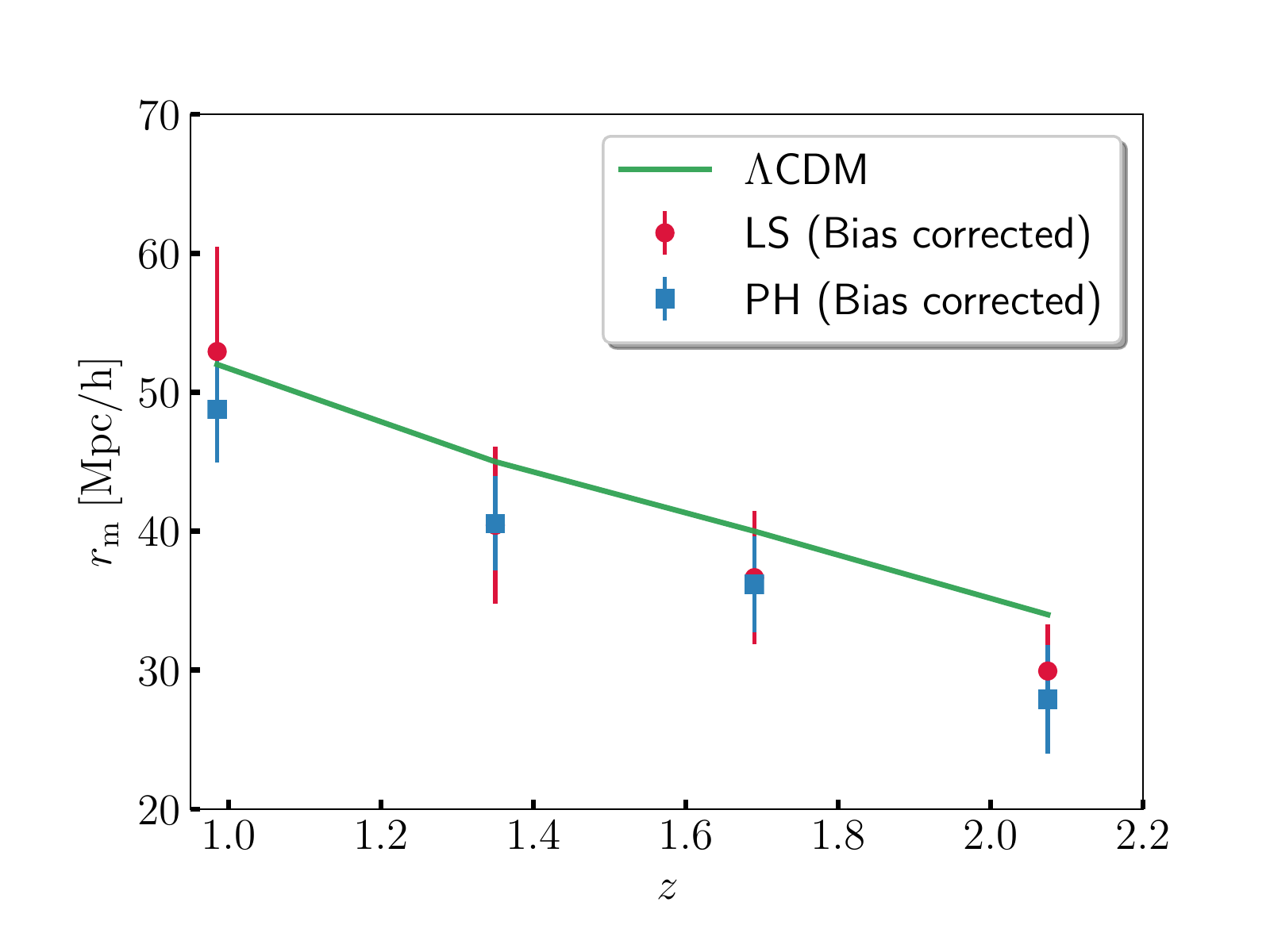}
\caption{Evolution of the homogeneity scale with redshift.} 
\label{fig2}
\end{figure}

The final measurements of both $r_{\rm hom}^{\rm m}$ and $r_{\rm hom}^{\rm th}$ are given in the second and third column of Tables~\ref{tableresultPH} and~\ref{tableresultLS} for the PH and LS estimators, respectively. The theoretical prediction of the homogeneity scale from our fiducial model is shown in the last column of both tables, as displayed in Fig.~\ref{fig2}. Clearly, $r_{\rm hom}^{\rm m}$ exhibits a decreasing trend for both estimators, and they are all fully compatible with the $r_{\rm hom}^{\rm th}$ values. Therefore, we conclude that there is a clear scale of cosmic homogeneity in the latest DR14 SDSS-IV quasar data, and that our results are well consistent with the $\Lambda$CDM paradigm. 

\section{Conclusion}

In this paper, we estimated the homogeneity scale from the SDSS-IV DR14 quasar sample. To our knowledge this is the first time that the transition to homogeneity  is probed in such large redshift range ($0.80 < z < 2.25$). We split the original data set in four redshift slices, whose mean redshifts are $\bar{z} = 0.985, 1.35, 1.690, 2.075$, and study a possible evolution of $r_{\rm hom}^{\rm m}$ with respect to the redshift. Our analyses were performed using the scaled counts-in-spheres approach, $\mathcal{N}(<r)$,  and its logarithmic derivative - the fractal correlation dimension $D_2$, using two different estimators, namely the Peebles-Hauser and Landy-Szalay, as defined by Eqs.~\ref{NPH} and~\ref{NLS}, respectively. Following previous analyses, we ascribed the scale of cosmic homogeneity $r_{\rm hom}$ to the characteristic where $D_2 \rightarrow 2.97$. The results we obtained are in good agreement with those recently discussed in the literature.

In order to test consistency between data and the SCM, we also estimated the bias of our quasar sample. We showed a clear evolution with redshift, which can be best described by a parameterisation of the type $b_{\rm q}(z) = b_0 + b_1(1+z)^2$. Considering the best-fit values of $b_0$ and $b_1$ for each estimator in $D_2$, we found that the scale of homogeneity of the matter distribution $r_{\rm hom}^{\rm m}$ indeed presents a decreasing trend with respect to the redshift, and that they are fully consistent with the homogeneity scale obtained from the theoretical matter power spectrum, $r_{\rm hom}^{\rm th}$, in all redshift ranges.  

Our results showed that there is a clear scale of cosmic homogeneity in the latest SDSS-IV quasar data, as predicted by the fundamental assumptions underlying the $\Lambda$CDM model. Further exploration of the variation of the homogeneity scale with respect to the redshift, as well as the impact of the clustering bias, angular selection function etc. on its estimate is currently in progress~\citep{goncalves18b}. As the amount of cosmological observations is expected to largely increase with the advent of next-generation surveys~\citep{lsst09, jpas14, ska15a, ska15b, euclid16}, we expect to improve our results, and thus establish the CP as a observationally valid physical assumption describing the observed Universe.  

\section*{Acknowledgments}

We acknowledge the use of the Sloan Sky Digital Survey data~\citep{york00}. RSG, GCC, and JCC thank CNPq for support. RSG acknowledges support from PNPD/CAPES. JSA acknowledges support from CNPq (Grants no. 310790/2014-0 and 400471/2014-0) and FAPERJ (grant no. 204282). CAPB acknowledges support from the South African SKA Project. The authors would also like to thank the anonymous referee for the useful suggestions that greatly improved the exposition of the paper.


\end{document}